\documentclass[12pt,a4paper]{article}
\input xy
\xyoption{all}
\usepackage{amsthm}
\usepackage{amsfonts,amsmath}
\usepackage{lscape}

\def\tr{\mbox{tr}}
\def\tp{\otimes} 
\def\es{\vspace{0.2cm}} 

\def\C{\mathbb{C}}

\newtheorem{thm}{Theorem}[section]
\newtheorem{prop}[thm]{Proposition}

\theoremstyle{definition}

\newtheorem{cor}[thm]{Corollary}

\theoremstyle{remark}

\title{Integrable boundary conditions for a non-abelian anyon chain with $D(D_3)$ symmetry}
\author{
K.A. Dancer, P.E. Finch, P.S. Isaac and J. Links
}

\begin{document}
\maketitle

\begin{abstract}
A general formulation of the Boundary Quantum Inverse Scattering Method is given which is applicable in cases where $R$-matrix solutions of the Yang--Baxter equation do not have the property of crossing unitarity. Suitably modified forms of the reflection equations are presented which permit the construction of a family of commuting transfer matrices. As an example, we apply the formalism to determine the most general solutions of the reflection equations for a solution of the Yang-Baxter equation with underlying symmetry given by the Drinfeld double $D(D_3)$ of the dihedral group $D_3$. This $R$-matrix does not have the crossing unitarity property. In this manner we derive integrable boundary conditions for an open chain model of interacting non-abelian anyons.
\end{abstract}

\section{Introduction}

The study of systems with non-abelian anyonic degrees of freedom currently attracts high interest, due to the possibilities for exploiting their topological properties to encode quantum information in a manner which is protected from decoherence \cite{nssfd2008}. An appropriate framework in which to formulate systems with anyonic symmetries is through the representation theory of quasi-triangular Hopf algberas \cite{d1986}, which includes the class of Drinfeld doubles of finite group algebras \cite{dpr91,Gould1993}. This latter class of algebras is particularly suited for the description of non-abelian anyons where the conjugacy classes
and centraliser subgroups of the finite group label generalised notions of the magnetic and electric
charges \cite{wild1998,kitaev2003}.  
Moreover, within the quasi-triangular Hopf algebra framework,  consistent braiding and fusion properties for anyonic theories are naturally obtained. The braiding properties are characterised by solutions of the Yang-Baxter equation {\it without} spectral parameter, which are realised through the universal $R$-matrix of the algebra. The fusion properties are given by  decompositions of tensor product representations of the Hopf algebra, which are governed by the coproduct structure. These fusion rules provide a means to construct interacting systems by assigning energies to the various possible multiplet structures. This then enables the study of one-dimensional (chain) models with local interactions, as has been recently undertaken in \cite{ftltkwf2007} using Fibonacci anyons. For this case the interaction energies were chosen in such a way that the local Hamiltonians provided representations of the Temperley-Lieb algebra, which necessarily means that the system is integrable and can be solved exactly. A study of a non-integrable non-abelian anyon chain can be found in \cite{by2007}.

The theory of integrable chains has a long history associated with the Quantum Inverse Scattering Method (QISM) \cite{STF1979}, which relies on a solution of the Yang--Baxter equation {\it with} spectral parameter to construct a family of commuting transfer matrices. The transfer matrix may be used to generate the conserved operators of an integrable quantum system. Following this procedure we have previously shown that, using the Drinfeld double $D(D_3)$ of the dihedral group $D_3$, there exists a spectral parameter dependent solution of the Yang-Baxter equation which can be used to construct an integrable interacting  non-abelian anyon chain \cite{DIL2006}. There the standard approach of the QISM was used, producing a chain with periodic boundary conditions. For open chain cases integrable boundary conditions are provided by solutions of the reflection equations, as was first elucidated by Sklyanin \cite{Sklyanin1988}, and is generally known as the Boundary Quantum Inverse Scattering Method (BQISM). Our goal is to extend the BQISM formalism in a manner which will enable the construction of a  non-abelian anyon open chain with integrable boundary conditions.   

In Sklyanin's original formulation of the BQISM several conditions were imposed on the $R$-matrix including $P$-symmetry, $T$-symmetry and crossing symmetry \cite{Sklyanin1988}. It was soon realised that the BQISM can be extended to cases where the $P$- and $T$-symmetry properties are relaxed to the more general $PT$-symmetry \cite{MezNep1991}, and the crossing symmetry property can be replaced by the more general crossing unitarity condition \cite{MN1992} (see equation (\ref{cu}) below for the definition). Later it was shown in \cite{LinksGould1996} that the BQISM can be formulated for cases without $PT$-symmetry. Here we further extend the formulation of the BQISM by removing the imposition of the crossing unitarity property.
This is necessary to construct integrable boundary conditions for the $D(D_3)$ anyon chain, as the $R$-matrix does not possess this property. 

In Section 2 we present the formulation of the BQISM for $R$-matrices without crossing unitarity. Using the explicit example provided by the $D(D_3)$ $R$-matrix of \cite{DIL2006}, in Section 3 we explicitly find the most general solutions of the reflection equations.  In Section 4 we use these results to derive a non-abelian  anyon chain with integrable boundary conditions, and concluding remarks are given in Section 5.

\section{BQISM for $R$-matrices without crossing unitarity}


Our first objective is to reformulate the BQISM with a minimum number of assumed properties imposed on the $R$-matrix solution of the Yang--Baxter equation. 
We start with invertible operators $R(z) \in \mbox{End }(V \tp V)$ and $L(z) \in \mbox{End }(V \tp W)$ which satisfy the Yang--Baxter equation on $\mbox{End }(V \tp V \tp V)$:
\begin{equation}
 R_{12}(xy^{-1}) R_{13}(x) R_{23}(y) = R_{23}(y) R_{13}(x) R_{12}(xy^{-1}),
\label{ybe}
\end{equation}
\noindent
and the intertwining relation on $\mbox{End }(V \tp V \tp W):$
\begin{equation} 
	R_{12}(xy^{-1}) L_{13}(x) L_{23}(y) = L_{23}(y) L_{13}(x) R_{12}(xy^{-1}). 
\label{rll}
\end{equation}
Here the subscripts indicate on which vector spaces each operator acts, so for example $R_{23}(z) = I \tp R(z).$ Solutions
 to the Yang--Baxter equation (\ref{ybe}) are referred to as  $R$-matrices, while $L(x)$ appearing in (\ref{rll}) is called an $L$-operator.
We impose only the following conditions on $R(z)$:
\begin{enumerate}
	\item $R^{t_{1}}(z)$ is invertible and
	\item $R(z)$ obeys regularity, i.e. $R(1)=P$.
\end{enumerate}
Here $t_{1}$ denotes the partial transpose over the first space and $P \in \mbox{End }(V \tp V)$ is the usual permutation operator defined by
$$ P(v \tp w) = w \tp v, \hspace{1cm} v,w \in V. $$

\noindent The following theorem is reproduced from \cite{Takizawa2004}:

\begin{thm}
If $R(z)$ is an $R$-matrix satisfying the regularity property then it also satisfies the unitarity property, i.e.
$$ R_{12}(z)R_{21}(z^{-1}) = f(z) I \tp I, $$
with $f(z)$ a scalar function satisfying $f(z) = f(z^{-1})$.
\label{meroeht}
\end{thm}

\begin{proof}
Let $R(z)$ be an $R$-matrix satisfying regularity. Then
\begin{alignat*}{2}
&&R_{12}(z)R_{13}(1)R_{23}(z^{-1}) \;&=\; R_{23}(z^{-1})R_{13}(1)R_{12}(z) \\
&\Rightarrow \quad   & R_{12}(z)P_{13}R_{23}(z^{-1}) \;&= \; R_{23}(z^{-1})P_{13}R_{12}(z) \\
&\Rightarrow \quad  & R_{12}(z)R_{21}(z^{-1})P_{13} \;&=\;  R_{23}(z^{-1})R_{32}(z)P_{13} \\
&\Rightarrow \quad  & R_{12}(z)R_{21}(z^{-1}) \;&=\;  R_{23}(z^{-1})R_{32}(z)
\end{alignat*}
The left hand side acts trivially on the the third space while the right hand side acts trivially on the first. Combining these it follows that $R_{12}(z)R_{21}(z^{-1})$ must be a scalar of the identity, and that the scalar function is invariant under $z \rightarrow z^{-1}$.
\end{proof}
\noindent
Utilising the condition that $R^{t_{1}}(z)$ is invertible we define the operator
\begin{equation} \label{DefCurlR}
  \mathcal{R}_{12}(z) =  [(R_{21}^{t_{1}}(z))^{-1}]^{t_{1}},
\end{equation}
which by definition implies
\begin{equation}
 \mathcal{R}^{t_{1}}_{12}(z) R_{21}^{t_{1}}(z) = \mathcal{R}_{21}^{t_{2}}(z) R_{12}^{t_{2}}(z) = 1. 
\label{unit}
\end{equation}

We now introduce two reflection equations
\begin{eqnarray}
  R_{12}(xy^{-1}) K_{1}^{-}(x) R_{21}(xy)K_{2}^{-}(y) & =\es & K_{2}^{-}(y)R_{12}(xy)K_{1}^{-}(x)R_{21}(xy^{-1}), \hspace{0.5cm}  \label{REMinus} \\
  R_{12}(yx^{-1}) K_{1}^{+}(x) \mathcal{R}_{21}(xy) K_{2}^{+}(y) & =\es & K_{2}^{+}(y) \mathcal{R}_{12}(xy) K_{1}^{+}(x) R_{21}(yx^{-1}), \label{REPlus} 
\end{eqnarray}
where $K^+(z), K^-(z) \in \mbox{End }(V)$ are known as reflection matrices.
The matrices $K^{+}(z)$, $K^{-}(z)$, $R(z)$ and $L(z)$ will enable us to construct an integrable model on an open chain. The transfer matrix is defined as
$$ t(z) = \tr_{a}\left[ K_{a}^{+}(z) T(z) \right], $$
where $\tr_{a}$ is the trace over space $a$ and $T(z)$ is the {\it double monodromy matrix}
$$ T(z) = L_{aN}(z)...L_{a1}(z)K^{-}_{a}(z)L_{a1}^{-1}(z^{-1})...L_{aN}^{-1}(z^{-1}). $$
It is known and easily verifiable from (\ref{rll}) and (\ref{REMinus}) that
\begin{equation} 
	R_{12}(xy^{-1}) T_{13}(x) R_{21}(xy)T_{23}(y) = T_{23}(y)R_{12}(xy)T_{13}(x)R_{21}(xy^{-1}).
\label{inter}
\end{equation}
\begin{prop}
The transfer matrices $t(x), t(y)$ commute for all $x, y \in \mathbb{C}$.
\end{prop}
\begin{proof}
\begin{eqnarray*}
 && f({x}{y^{-1}}) t(x)t(y) \\
 &&\qquad  = \left(\tr_{1}\tp \tr _{2}\right) \left\{ f({x}{y^{-1}}) K_{2}^{+}(y)K_{1}^{+}(x)^{t_{1}} T_{13}^{t_{1}}(x) T_{23}(y) \right\} \\
&&\qquad  = \left(\tr_{1}\tp \tr _{2}\right) \left\{ f({x}{y^{-1}}) K_{2}^{+}(y)K_{1}^{+}(x)^{t_{1}} \mathcal{R}^{t_{1}}_{12}(xy) R_{21}^{t_{1}}(xy) T_{13}^{t_{1}}(x) T_{23}(y) \right\} \\
&& \qquad = \left(\tr_{1}\tp \tr _{2}\right) \left\{ f({x}{y^{-1}}) K_{2}^{+}(y) \mathcal{R}_{12}(xy) K_{1}^{+}(x) T_{13}(x) R_{21}(xy) T_{23}(y) \right\} \\
&&\qquad =\left(\tr_{1}\tp \tr _{2}\right) \left\{ \left[K_{2}^{+}(y) \mathcal{R}_{12}(xy) K_{1}^{+}(x) R_{21}({y}{x^{-1}}) \right] \left[R_{12}({x}{y^{-1}}) T_{13}(x) R_{21}(xy)T_{23}(y) \right] \right\} \\
  &&\qquad = \left(\tr_{1}\tp \tr _{2}\right) \left\{ \left[R_{12}({y}{x^{-1}}) K_{1}^{+}(x) \mathcal{R}_{21}(xy) K_{2}^{+}(y) \right] \left[T_{23}(y)R_{12}(xy)T_{13}(x)R_{21}({x}{y^{-1}}) \right] \right\} \\
 &&\qquad = \left(\tr_{1}\tp \tr _{2}\right) \left\{ f({x}{y^{-1}}) K_{1}^{+}(x) \mathcal{R}_{21}(xy) K_{2}^{+}(y) T_{23}(y) R_{12}(xy) T_{13}(x)  \right\} \\
  &&\qquad =\left(\tr_{1}\tp \tr _{2}\right) \left\{ f({x}{y^{-1}}) K_{1}^{+}(x) K_{2}^{+}(y)^{t_{2}} \mathcal{R}_{21}^{t_{2}}(xy) R_{12}^{t_{2}}(xy) T_{23}^{t_{2}}(y) T_{13}(x) \right\} \\
  &&\qquad = \left(\tr_{1}\tp \tr _{2}\right) \left\{ f({x}{y^{-1}}) K_{1}^{+}(x) K_{2}^{+}(y)^{t_{2}} T_{23}^{t_{2}}(y) T_{13}(x) \right\} \\
  &&\qquad = f(xy^{-1}) t(y) t(x)
\end{eqnarray*}
where we have used equations (\ref{unit},\ref{REPlus},\ref{inter}) and Theorem \ref{meroeht}.
\end{proof}

We now impose the limit condition 
$$ K^{-}(1) = I $$
and only consider instances for  which $\tr\left(K^{+}(1)\right) \neq 0$.
Then in the case $L(z) = R(z)$ the global  Hamiltonian on an open chain with boundary fields is defined in the following way:
\begin{eqnarray}
	\mathcal{H} 
  & =\es & \frac{c}{2 \tr\left(K^{+}(1)\right)} \left[\left. \frac{d}{dz}\left(t(z)\right) \right|_{z=1} - \left. \tr\left(  \frac{d}{dz}\left(K^{+}(z)\right)\right)\right|_{z=1} \right] \nonumber \\
  & =\es & \sum_{i=1}^{N-1} H_{i,i+1} + \frac{c}{2}\left. \frac{d}{dz}\left(K^{-}_{1}(z)\right) \right|_{z=1} + \frac{\tr_{a}\left( c K^{+}_{a}(1) H_{N,a} \right)  }{\tr\left(K^{+}(1)\right)},
\label{ham}
\end{eqnarray}
where $c\in{\mathbb C}$ and the local Hamiltonians are given by
$$ H_{i,i+1} =c\, P_{i,i+1} \left. \frac{d}{dz}\left(R(z)_{i,i+1}\right) \right|_{z=1}. $$
We will refer to the first term of (\ref{ham}) as the bulk Hamiltonian, while the second and third terms describe boundary field interactions. By construction the global Hamiltonian (\ref{ham}) commutes with the transfer matrix $t(z)$, which means that the Hamiltonian is necessarily integrable. The conserved operators commuting with the Hamiltonian are obtained as the co-efficient operators in the series expansion of $t(z)$.

\section{Reflection matrices for an $R$-matrix associated with $D(D_{3})$}
We now apply the above formalism to solve for the reflection matrices satisfying equations (\ref{REMinus}) and (\ref{REPlus}). 
This $R$-matrix we use is constructed from the representation theory of $D(D_{3})$ \cite{DIL2006}, with associated $L$-operators given in \cite{DL2008}. Explicitly, we have
\begin{equation}  R(z) = 
  \left( \begin {array}{ccccccccc}
  1&0&0&0&0&0&0&0&0  \\
  0&0& \frac{z( z-1)}{z^{2}-z+1} & \frac{z}{z^{2}-z+1}&0&0&0&\frac{1-z}{z^{2}-z+1}&0 \\
  0& \frac{z( z-1)}{z^{2}-z+1} &0&0&0&\frac{1-z}{z^{2}-z+1}& \frac{z}{z^{2}-z+1} &0&0 \\
  0& \frac{z}{z^{2}-z+1} &0&0&0& \frac{z( z-1)}{z^{2}-z+1} &\frac{1-z}{z^2-z+1}&0&0 \\
  0&0&0&0&1&0&0&0&0 \\
  0&0&\frac{1-z}{z^{2}-z+1}& \frac{z( z-1)}{z^{2}-z+1} &0&0&0&\frac{z}{z^{2}-z+1}&0 \\
  0&0&\frac{z}{z^{2}-z+1}&\frac{1-z}{z^{2}-z+1}&0&0&0& \frac{z( z-1)}{z^{2}-z+1} &0 \\
  0&\frac{1-z}{z^{2}-z+1}&0&0&0& \frac{z}{z^{2}-z+1} &\frac{z( z-1)}{z^{2}-z+1} &0&0 \\
  0&0&0&0&0&0&0&0&1
\end {array} \right). 
\label{rm}
\end{equation} 
The properties of the $R$-matrix include regularity, and consequently unitarity.  However it can be verified that the $R$-matrix does not satisfy the crossing unitarity condition, i.e. there does not exist $M\in{\rm End}(V)$ and $\lambda\in{\mathbb C}$ such that  
\begin{eqnarray}
R_{12}^{t_1}(\lambda z)M_1R_{21}^{t_1}(z^{-1})M_1^{-1}=f(z) I \otimes I. 
\label{cu}
\end{eqnarray} 
This is in contrast to $R$-matrices obtained from loop representations of affine quantum algebras, for which equation (\ref{cu}) is always satisfied \cite{rsts1990}. 

We first calculate $\mathcal{R}(z)$ defined by equation (\ref{DefCurlR}):
$$  \mathcal{R}(z) = \frac{(z^2-z+1)}{(z-1)(z^{3}-1)}
  \left( \begin {array}{ccccccccc}
  {z}^{2}+1&0&0&0&z&0&0&0&z \\
  0&0&1&-z&0&0&0&{z}^{2}&0 \\
  0&1&0&0&0&{z}^{2}&-z&0&0 \\
  0&-z&0&0&0&1&{z}^{2}&0&0 \\
  z&0&0&0&{z}^{2}+1&0&0&0&z \\
  0&0&{z}^{2}&1&0&0&0&-z&0 \\
  0&0&-z&{z}^{2}&0&0&0&1&0 \\
  0&{z}^{2}&0&0&0&-z&1&0&0 \\
  z&0&0&0&z&0&0&0&{z}^{2}+1
\end {array} \right). $$ 
These are the two operators required to construct the reflection matrices $K^{-}(z)$ and $K^{+}(z)$. 

\subsection{Special case of the reflection equation}
To determine the possible matrices, $K^{-}(z)$ and $K^{+}(z)$, which satisfy equations (\ref{REMinus}) and (\ref{REPlus}) it is first convenient to determine all non-diagonal invertible matrices $K(z)$ which satisfy the equation
\begin{equation} \label{RESpecial}
	K_{2}(y) \check{R}_{12}(0) K_{2}(z_{0}) \check{R}_{12}(0) = \check{R}_{12}(0) K_{2}(z_{0}) \check{R}_{12}(0) K_{2}(y),
\end{equation}
where $y, z_{0} \in \C$, $z_{0}$ is fixed and $\check{R}(0) = PR(0)$. We scale $K(z)$ so that the entries of $\lim_{z \rightarrow z_{0}} K(z)$ are all finite and that at least one is non-zero, as is always possible. 
Throughout this section we write $K(z)$ in the form
$$ K(z) = \sum_{i,j=1}^{3} h_{i,j}(z) E^i_{j} $$
where $E^i_{j}$ denotes the elementary matrix with a 1 in the $i$th row and $j$th column. We consider the indices of the functions $h_{i,j}(z)$ and elementary matrices $E^i_{j}$ modulo 3. Using this notation, $K(z)$ is a solution to 
equation (\ref{RESpecial}) if and only if 
$$ h_{i,j}(z_{0})h_{k,l}(y) = h_{i,j+k+2l}(z_{0}) h_{2i+2k,2i+2l}(y) $$
for all $ 1 \leq i,j,k,l \leq 3 $. 

\begin{prop} \label{prop1}
If $K(z)$ satisfies equation $(\ref{RESpecial})$ and $h_{a,a}(z_{0}) = 0$ for some $1 \leq a \leq 3$ then $h_{a,j}(z_{0}) = 0$ for all $1 \leq j \leq 3$.
\end{prop}
\begin{proof}
Assume there is an integer $b$ such that  $h_{a,b}(z_{0}) \neq 0$. As $K(z)$ satisfies equation $(\ref{RESpecial})$, we have
$$ h_{a,a}(z_{0})h_{a,l}(y) = h_{a,2a+2l}(z_{0}) h_{a,2a+2l}(y) \hspace{0.5cm}\Rightarrow\hspace{0.5cm} h_{a,b}(z_{0}) h_{a,b}(y) = 0.$$
This contradiction proves the proposition.
\end{proof}

\begin{cor} \label{cor1}
If $K(z)$ is a solution to equation (\ref{RESpecial}) then $K(z_0)$ has at least one non-zero diagonal entry.
\end{cor}
\begin{proof}
Assume there is a solution where all the diagonal entries of $K(z_0)$ are zero. It follows from Proposition 
\ref{prop1} that $h_{a,j}(z_{0}) = 0$ for all $1 \leq a,j \leq 3$, which contradicts our 
requirement that at least one entry of $K(z_{0})$ is non-zero.
\end{proof}

\begin{prop} \label{prop2}
If $K(z)$ is a non-diagonal matrix satisfying equation (\ref{RESpecial}) and $h_{a,a}(z_{0}) \neq 0$ for some  $1 \leq a \leq 3$ then $h_{a,a+b}(z_{0}) \neq 0$ for $b \in \{1,2\}$. Furthermore, $h_{a,a+2}(z) = \alpha h_{a,a+1}(z)$ and $\alpha^{3} = 1$ where $ \alpha  = {h_{a,a+1}(z_{0})}/{h_{a,a}(z_{0})}$.
\end{prop}
\begin{proof}
We now assume that there exists a $b \in \{1,2\} $ such that $h_{a,a+b}(z_{0}) = 0$. This leads to 
$$ h_{a,a+b}(z_{0})h_{a,a+b}(y) = h_{a,a}(z_{0}) h_{a,a+2b}(y)  \hspace{0.5cm}  \Rightarrow  \hspace{0.5cm}  h_{a,a+2b}(y) = 0. $$
Therefore $h_{a,a+2b}(y) = 0$, which implies through the same argument that $h_{a,a+b}(y) = 0$. Hence 
$$ h_{a,a+1}(y) = h_{a,a+2}(y) = 0. $$
But 

$$h_{a,a}(z_{0}) h_{l+c,l}(y) = h_{a,a+c}(z_{0})h_{2a+2l+2c,2a+2l}(y) \hspace{0.5cm}  \Rightarrow  \hspace{0.5cm}  h_{l+c,l}(y) = 0$$
for $1 \leq l \leq 3$ and $c\in\{1,2\}$. This implies that if $h_{a,a+b}(z_{0}) = 0$ for some integer $b\in \{1,2\}$ then $K(z)$ is diagonal, which is a contradiction. Hence $h_{a,a+b}(z_{0}) \neq 0$ for $b \in \{1,2\}$. \\

To show the other half of the proposition we set $ \alpha = {h_{a,a+1}(z_{0})}/{h_{a,a}(z_{0})}$. Then 
$$ \begin{array}{rcl}
  h_{a,a+1}(z_{0})h_{a,a+1}(y) = h_{a,a}(z_{0}) h_{a,a+2}(y)  & \hspace{0.2cm} \Rightarrow \hspace{0.2cm} &  h_{a,a+2}(y) = \alpha h_{a,a+1}(y) \hspace{0.7cm} \mbox{and} \\
 h_{a,a+2}(z_{0})h_{a,a+2}(y) = h_{a,a}(z_{0}) h_{a,a+1}(y)  &  \Rightarrow  &  [\alpha^{3}-1]h_{a,a+1}(y) = 0 . 
\end{array} $$
This completes the proof.
\end{proof}

\begin{prop} \label{prop3}
For a non-diagonal matrix, $K(z)$, which satisfies equation (\ref{RESpecial}) and has $h_{a,a}(z_{0}) \neq 0$ and $h_{a+1,a+1}(z_{0}) \neq 0$ for some $1 \leq a \leq 3$ then $h_{a+2,a+2}(z_{0}) \neq 0$.
\end{prop}
\begin{proof}
We find
$$ h_{a+1,a+1}(z_{0})h_{a+2,a+2}(y) = h_{a+1,a+1}(z_{0}) h_{a,a}(y) \hspace{0.5cm}  \Rightarrow  \hspace{0.5cm} h_{a+2,a+2}(y) = h_{a,a}(y), $$
which proves the proposition.
\end{proof}

\begin{prop} \label{prop4}
Any non-diagonal matrix, $K(z)$, which satisfies equation (\ref{RESpecial}) and has $h_{i,i}(z_0) \neq 0$ for all $1 \leq i \leq 3$ is of the form
$$ K(z) = 
\left( \begin{array}{ccc}
  A(z) & \alpha B(z) & \alpha^{2} B(z) \\
  \beta^{2} B(z) & A(z) & \beta B(z) \\
  \gamma B(z) & \gamma^{2} B(z) & A(z) \\
\end{array} \right) $$
where $\alpha^{3} = \beta^{3} = \gamma^{3} = \alpha\beta\gamma = 1$ and $A(z_{0}) = B(z_{0}) = 1$.
\end{prop}
\begin{proof}
As $h_{1,1}(z_{0}) \neq 0$ and we are free to scale by a constant we set $h_{1,1}(z_{0}) = 1 $. We find that
$$ h_{i,i}(z_{0}) [h_{l,l}(y) -h_{2i+2l,2i+2l}(y)] =  0  \hspace{0.5cm}  \Rightarrow  \hspace{0.5cm} h_{l,l}(y) = h_{j,j}(y) $$
for all $1 \leq i,j,l \leq 3$. \\

We now define the following variables
$$ \lambda_{a}  = h_{a,a+1}(z_{0}), $$
for $1 \leq a \leq 3$. We see that from the previous proposition that $\lambda_{a}^{3} = 1$ and 
$$ h_{a,a+2}(y) = \lambda_{a} h_{a,a+1}(y) $$
for all $a \in \{1,2,3\}$. This gives the properties required for within each row, 
but we still need to relate the entries down each column. We have that
$$ h_{a,a+c}(z_{0})h_{a+2c,a+c}(y) = h_{a,a+2c}(z_{0}) h_{a+c,a+2c}(y) $$
for $a \in \{1,2,3\}$ and $c \in \{1,2\}$. Hence
$$ h_{a+2,a+1}(y) = \lambda_{a} h_{a+1,a+2}(y) \hspace{0.5cm}  \Rightarrow  \hspace{0.5cm} 
\lambda_{a+2} h_{a+2,a}(y) = \lambda_{a} h_{a+1,a+2}(y) $$
Expressing this explicitly we have that
$$ \lambda_{1}\lambda_{3} h_{1,2}(y) = \lambda_{2} \lambda_{1}h_{2,3}(y) = \lambda_{2} \lambda_{3} h_{3,1}(y) $$
Hence the off-diagonal entries are all scalar multiples of each other. Moreover, by considering the above equation 
at $y = z_0$ we find
$$ \lambda_{1}^2\lambda_{3} = \lambda_{1}\lambda_{2}^{2} \hspace{0.5cm}  \Rightarrow  \hspace{0.5cm} \lambda_{1}\lambda_{3}\lambda_{2} =1. $$
This proves the proposition.
\end{proof}

\begin{prop} \label{prop5}
If $K(z)$ satisfies equation (\ref{RESpecial}) and $K(z_{0})$ has only one non-zero diagonal entry  then $K(z)$ can be 
written in the form (after basis transformation and scaling)
$$ K(z) = 
\left( \begin{array}{ccc}
  A(z) & \alpha B(z) & \alpha^{2} B(z) \\
   \alpha D(z) & C(z) & E(z) \\
  D(z) & \alpha E(z) & C(z) \\
\end{array} \right) $$
where $\alpha^{3} = 1$, $A(z_{0}) = B(z_{0}) = 1$ and $C(z_{0}) = D(z_{0}) = E(z_{0}) = 0$.
\end{prop}
\begin{proof}
First note that $\check{R}$ is invariant under relabelling of the indices, so without loss of generality we can 
set  $h_{1,1}(z_{0}) = 1$. We set $A(z) = h_{1,1}(z)$, so $A(z_0) = 1$.  By Proposition \ref{prop2} we know that $h_{1,2}(z_{0}) \neq 0$, and 
thus we set $h_{1,2}(z) = \alpha B(z)$ with $B(z_{0}) = 1$. By Proposition \ref{prop2} we also have that $\alpha^{3} = 1$ and $h_{1,3}(z) = \alpha^{2} B(z)$. \\

\noindent We now consider the diagonal entries and see that
$$ h_{1,1}(z_{0})h_{3,3}(y) = h_{1,1}(z_{0}) h_{2,2}(y). $$
It follows that $h_{3,3}(z) = h_{2,2}(z)$. We let $C(z) = h_{2,2}(z)$. \\

\noindent We now use
$$ h_{1,1}(z_{0})h_{2,l}(y) = h_{1,2l}(z_{0}) h_{3,2+2l}(y). $$
This gives
$$ h_{2,1}(y) = \alpha h_{3,1}(y) \mbox{   and   } h_{2,3}(y) = \alpha^{2} h_{3,2}(y). $$
We let $D(z) = h_{3,1}(z)$ and $E(z) = h_{2,3}(z)$. Lastly, note that $C(z_{0}) = 0$ and hence by Proposition \ref{prop1}
we have $D(z_{0}) = E(z_{0}) = 0$. 
\end{proof}

We have classified all the possible non-diagonal matrix solutions to equation (\ref{RESpecial}). These can be classified by the number of non-zero diagonal elements of $K(z_{0})$. In Corollary \ref{cor1} and Proposition \ref{prop3} it was shown that $K(z)$ can only have one or three non-zero diagonal elements. Moreover, these solutions can all be written (after basis transformation and scaling) in the forms given in Propositions \ref{prop4} and \ref{prop5} respectively.

\subsection{Reflection matrix $K^{-}(z)$}
We now turn our attention to the reflection equation (\ref{REMinus}).  Setting $x = \infty$ and using regularity, we obtain the equation
$$ R_{21}(0) K_{1}^{-}(\infty) R_{12}(0)K_{2}^{-}(y)  = K_{2}^{-}(y)R_{21}(0)K_{1}^{-}(\infty)R_{12}(0). $$
This is equivalent to equation (\ref{RESpecial}) with $z_{0}=\infty$. Using the previous section we know all the possible forms of $K^{-}(z)$. We now only need check under which conditions, if any, these forms satisfy the reflection equation \eqref{REMinus}.

\begin{prop} \label{prop6}
The only diagonal matrices satisfying equation (\ref{REMinus}) are scalars of the identity.
\end{prop}
\begin{proof}
The proof involves a straightforward calculation so we omit the details.
\end{proof}

\begin{prop} \label{prop7}
There are no invertible matrices satisfying equation (\ref{REMinus}) of the form
$$ K^{-}(z) = 
\left( \begin{array}{ccc}
  A(z) & \alpha B(z) & \alpha^{2} B(z) \\
  \alpha D(z) & C(z) & E(z) \\
  D(z) &   \alpha E(z) & C(z) \\
\end{array} \right) $$
where $\alpha^{3} = 1$, $A(\infty) = B(\infty) = 1$ and $C(\infty) = D(\infty) = E(\infty) = 0$.
\end{prop}
\begin{proof}
We first assume that such a $K^{-}(z)$ does exist. Substituting $K^{-}(z)$ into the reflection equation, we find it must 
satisfy the constraints
$$ D(y)B(x) = B(y)D(x) \hspace{0.5cm} \mbox{and} \hspace{0.5cm} E(y)B(x) = B(y)E(x). $$
This implies that $D(z)$ and $E(z)$ are both scalars of $B(z)$. Using the boundary conditions at $z= \infty$ we deduce that
$$ D(z) = E(z) = 0. $$
After imposing $D(z)=E(z)=0$ we obtain the second constraint equation
$$ B(x)B(y) = 0. $$
This is a contradiction, as $B(\infty) = 1$, and hence there are no solutions $K^-(z)$ of the above form.
\end{proof}

\begin{prop} \label{prop8}
All matrices satisfying equation (\ref{REMinus}) are either scalars of the identity or of the form 
$$ K^{-}(z) = 
\left( \begin{array}{ccc}
  w^2 + bz - z^2 & (1-z^{2}) & (1-z^{2}) \\
  w^{2} (1-z^{2}) & w^2 + bz - z^2 & w (1-z^{2}) \\
  w^{2} (1-z^{2}) & w (1-z^{2}) & w^2 + bz - z^2 \\
\end{array} \right), $$
for some $b \in \C$ and $w$ a cube root of unity, up to scaling and a change of basis.
\end{prop}
\begin{proof}
Let $K^{-}(z)$ be a solution to equation \eqref{REMinus} which is not a scalar of the identity. Then by 
the earlier propositions we know $K^{-}(z)$ can be written in the form
$$ K^{-}(z) = 
\left( \begin{array}{ccc}
  A(z) & \alpha B(z) & \alpha^{2} B(z) \\
  \beta^{2} B(z) & A(z) & \beta B(z) \\
  \gamma B(z) & \gamma^{2} B(z) & A(z) \\
\end{array} \right) $$
where $\alpha^{3} = \beta^{3} = \gamma^{3} = \alpha\beta\gamma = 1$ and $A(\infty) = B(\infty) = 1$. Substituting this into 
equation \eqref{REMinus}, we obtain the constraint equation
$$ \begin{array}{rcl}
  0 
  & =\es & -x(x^2 +\alpha^2\beta x^2 +\alpha\beta^{2}) - x(x^2+ \alpha\beta^2 +\alpha^2\beta)y^{2}\\
  &  \es & + (\alpha \beta^2 x^2 + \beta^2\alpha +x^2 +x^4 +2 \alpha^2\beta x^2)y \\
  &  \es & - y(1-x^2)^{2}f(x) + x(x^{2}-1)(y^{2}-1)f(y)
\end{array} $$
where $f(z) = {A(z)}/{B(z)}$. As the coefficient of $f(x)$ is linear in $y$, we take the double derivative with respect to $y$, obtaining
$$ 2\frac{(x^2+ \alpha\beta^2 +\alpha^2\beta)}{(x^2-1)}= 2f(y) + 4yf'(y) + (y^{2}-1)f(y). $$
The right hand side is independent of $x$, so we deduce that
$$ 0 = 1 + \alpha\beta^{2} + \alpha^{2}\beta = 1 + \alpha^{2}\beta + (\alpha^{2}\beta)^{2}. $$
This implies that $\alpha^{2}\beta$ is a primitive cube root of unity. We set 
$$ \beta = w \alpha, $$
where $w$ is a primitive cube root of unity. Our matrix becomes
$$ K^{-}(z) = 
\left( \begin{array}{ccc}
  A(z) & \alpha B(z) & \alpha^{2} B(z) \\
  w^{2}\alpha^{2} B(z) & A(z) & w \alpha B(z) \\
  w^{2} \alpha B(z) & w \alpha^{2} B(z) & A(z) \\
\end{array} \right). $$
The different choices of $\alpha$ are equivalent up to a basis transformation under which $\check{R}(z)$ is invariant, 
so without loss of generality we choose $\alpha =1$.  Now the differential equation reduces to 
$$ 2 = 2f(y) + 4yf'(y) + [y^2-1]f''(y). $$
This has the general solution
$$ f(y) = \frac{a + by - y^2}{1-y^2}, $$
where $a,b \in \C$. Without loss of generality we set
$$ A(z) = 1 \hspace{0.5cm} \mbox{and} \hspace{0.5cm} B(z) = \frac{1-z^2}{a+bz-z^2}. $$
Substituting $K^{-}(z)$ into equation (\ref{REMinus}), we find that we must have $a = w^{2}$. Furthermore if $a=w^2$ then $K^{-}(z)$ satisfies equation (\ref{REMinus}) for all $b \in \C$.
\end{proof}


\subsection{Reflection matrix $K^{+}(z)$}
In this section we construct solutions to the other reflection equation \eqref{REPlus}.  
Setting $x=0$ and using unitarity, we obtain the equation
$$ R_{21}(0) K_{1}^{+}(0) \mathcal{R}_{21}(0) K_{2}^{+}(y) = K_{2}^{+}(y) \mathcal{R}_{12}(0) K_{1}^{+}(0) R_{12}(0), $$
This is equivalent to equation (\ref{RESpecial}) with $z_{0}=0$ as $\mathcal{R}_{21}(0) = R_{12}(0)$. Using the previous section we know all the possible forms of $K^{+}(z)$. It remains to check whether these forms provide solutions 
to the reflection equation \eqref{REPlus}, and if so, under what conditions.

\begin{prop} \label{prop9}
The only diagonal reflection matrices satisfying equation (\ref{REPlus}) are scalars of the identity.
\end{prop}
\begin{proof}
The proof involves a straightforward calculation so we omit the details.
\end{proof}

\begin{prop} \label{prop10}
There are no invertible matrices satisfying equation (\ref{REPlus}) of the form
$$ K^{+}(z) = 
\left( \begin{array}{ccc}
  A(z) & \alpha B(z) & \alpha^{2} B(z) \\
   \alpha D(z) & C(z) & E(z) \\
  D(z) &   \alpha E(z) & C(z) \\
\end{array} \right) $$
where $\alpha^{3} = 1$, $A(0) = B(0) = 1$ and $C(0) = D(0) = E(0) = 0$.
\end{prop}
\begin{proof}
Directly substituting $K^{+}(z)$ into equation (\ref{REPlus}) we find the constraints
$$ D(y)B(x) = B(y)D(x). $$
This implies that $D(z)$ is a scalar of $B(z)$. As $D(0) = 0$ we deduce that
$$ D(z) = 0. $$
We also obtain the constraint equations.
$$ C(y)A(x) = A(y)C(x) \hspace{0.5cm} \mbox{and} \hspace{0.5cm} E(y)B(x) = B(y)E(x) $$
Using similar reasoning, we find that
$$ E(z) = C(z) = 0. $$
Hence $K^+(z)$ is not invertible, which proves the proposition.
\end{proof}

\begin{prop} \label{prop11}
All matrices satisfying equation (\ref{REPlus}) are either scalars of the identity or scalars of 
$$ K^{+}(z) = 
\left( \begin{array}{ccc}
  1 + bz - wz^2 & (1 - w^{2}z^{2}) & (1 - w^{2}z^{2}) \\
  w^{2} (1 - w^{2}z^{2}) & 1 + bz - wz^2 & w (1 - w^{2}z^{2}) \\
  w^{2} (1 - w^{2}z^{2}) & w (1 - w^{2}z^{2}) & 1 + bz - wz^2 \\
\end{array} \right), $$
for some $b \in \C$ and $w$ a primitive cube root of unity.
\end{prop}
\begin{proof}
By 
the earlier propositions, if $K^{+}(z)$ is not a scalar of the identity then it must be of the form
$$ K^{+}(z) = 
\left( \begin{array}{ccc}
  A(z) & \alpha B(z) & \alpha^{2} B(z) \\
  \beta^{2} B(z) & A(z) & \beta B(z) \\
  \gamma B(z) & \gamma^{2} B(z) & A(z) \\
\end{array} \right) $$
where $\alpha^{3} = \beta^{3} = \gamma^{3} = \alpha\beta\gamma = 1$ and $A(0)=B(0)=1$. Using a similar technique to that used to prove Proposition \ref{prop8}, we find a constraint equation and differentiate twice with respect to $x$. 
We obtain the following two equations:

$$ \begin{array}{rcl}
	2\alpha \beta^2 & = & 2 f(y) + 4y f'(y) + [y^2 + 1 + \alpha\beta^{2}]f''(y) \\
	0 & = & (1 + \beta\alpha^2 + (\beta\alpha^2)^2)\left[2f'(y) +yf''(y) \right]
\end{array} $$
where $f(y) = {A(y)}/{B(y)}$. 
These equations only have a solution in common when 
$$ (1 + \beta\alpha^2 + (\beta\alpha^2)^2) = 0. $$
Hence we must have $\beta = w\alpha = w^{2} \gamma$ where $w$ is a primitive cube root of unity. Furthermore, the three different choices of $\alpha$ are equivalent up to a basis transformation under which $R(z)$ and $\mathcal{R}(z)$ are both invariant, so without loss of generality we can choose $\alpha = 1$. The differential equation simplifies to
$$ 2w^2 = 2 f(y) + 4y f'(y) + [y^2 - w]f''(y). $$
This has the general solution
$$ f(z) = \frac{a + bz - wz^2}{1 - w^{2}z^{2}} $$
where $a,b \in \mathbb{C}$. However $f(0) = {A(0)}/{B(0)} = 1$, so $a=1$. We set
$$ A(z) = 1 + bz - wz^2 \hspace{0.5cm} \mbox{and} \hspace{0.5cm} B(z) = 1 - w^{2}z^{2}. $$ 
Substituting this into the reflection equation \eqref{REPlus}, we find this is a solution for all $b\in \mathbb{C}$. 
This completes the proof.
\end{proof}

\section{An integrable  Hamiltonian with open boundary conditions}
We found in the previous section all possible solutions $K^{+}(z)$ and $K^{-}(z)$ of the reflection equations \eqref{REMinus} and \eqref{REPlus} for our $R$-matrix with non-abelian anyonic symmetry. To calculate the global Hamiltonian with non-trivial boundary terms, we use the reflection matrices in the following form:
$$ K^{-}(z) = 
\left( \begin{array}{ccc}
  1 & \alpha \frac{(1-z^{2})}{w^2 + az - z^2} & \alpha^{2} \frac{(1-z^{2})}{w^2 + az - z^2} \\
  \alpha^{2}  w^{2} \frac{(1-z^{2})}{w^2 + az - z^2} & 1 & \alpha  w \frac{(1-z^{2})}{w^2 + az - z^2} \\
  \alpha  w^{2} \frac{(1-z^{2})}{w^2 + az - z^2} & \alpha^{2} w \frac{(1-z^{2})}{w^2 + az - z^2} & 1 \\
\end{array} \right), $$
and
$$ K^{+}(z) = 
\left( \begin{array}{ccc}
  1 + bz - w^{j}z^2 & \beta(1 - w^{2j}z^{2}) & \beta^{2} (1 - w^{2j}z^{2}) \\
  w^{2j}\beta^{2}(1 - w^{2j}z^{2}) & 1 + bz - w^{j}z^2 & w^{j} \beta(1 - w^{2j}z^{2}) \\
  w^{2j} \beta(1 - w^{2j}z^{2}) & w^{j}\beta^{2}(1 - w^{2j}z^{2}) & 1 + bz - w^{j}z^2 \\
\end{array} \right), $$
where $\alpha,\beta \in \{1,w,w^2\}$, $j \in \{1,2\}$, $a,b\in \C$ and $w$ is a primitive cube root of unity. As we are only considering the case where $\tr(K^{+}(1)) \neq 0$ we impose that $b \neq w^{j} -1$. Choosing $c=i$ in (\ref{ham}) yields the local Hamiltonian as given in \cite{DIL2006}:

$$ H_{i,i+1}=\sum_{\gamma \in D_3}i (E^{\gamma(1)}_{\gamma(2)} \tp E^{\gamma(2)}_{\gamma(3)} 
	- E^{\gamma(2)}_{\gamma(3)} \tp E^{\gamma(1)}_{\gamma(2)} )$$

\noindent where the elements $\gamma\in D_3$ are written as permutations of $\{1,2,3\}$. (Recall that $D_3$ is isomorphic to the permutation group $S_3$.) 
The bulk Hamiltonian, which describes nearest neighbour interactions between anyons with local three-dimensional state spaces, commutes with the action of $D(D_3)$ algebra \cite{DIL2006}.
For the boundary terms we find 
$$ \frac{i}{2}\left. \frac{d}{dz}\left(K^{-}_{1}(z)\right) \right|_{z=1} = A
\left( \begin{array}{ccc}
0 & \alpha & \alpha^{2} \\
\alpha^{2}w^{2} & 0 & \alpha w \\
\alpha w^{2} & \alpha^{2}w & 0
\end{array} \right) $$
and 
$$ \frac{\tr_{a}\left( i K^{+}_{a}(1) H_{N,a} \right)  }{\tr\left(K^{+}(1)\right)} = 
B\left( \begin{array}{ccc}
0 & \beta & \beta^{2} \\
\beta^{2}w^{j} & 0 & \beta w^{2j} \\
\beta w^{j} & \beta^{2}w^{2j} & 0
\end{array} \right) $$
where $A = i({1-w^{2}+a})^{-1} \in \C$ and $B = -i({1-w^j+b})^{-1} \in \C$. We choose $a=i\omega X^{-1} +\omega^{-1} -1 ,\, b=w^j-1-i\omega^{-j}Y^{-1}$ with $X,\,Y \in{\mathbb R}$ unconstrained parameters. This leads to $A = w^{-1}X, B =w^jY$, resulting in a global Hamiltonian which is hermitian.

Alternatively, one can choose the reflection matrices to be $K^+(z)=K^-(z)=I$. This yields an open chain Hamiltonian without boundary interaction terms. It is also possible to have one end of the chain with a boundary interaction term, while the other end is without a boundary interaction term.

\section{Summary} 

We reformulated the BQISM through a pair of reflection equations in a fashion which does not rely on the $R$-matrix solution of the Yang--Baxter equation satisfying the crossing unitarity condition. This was motivated by the case of the $R$-matrix (\ref{rm}) associated with the quasi-triangular Hopf algebra $D(D_3)$, for which crossing unitarity does not hold. We then proceeded to determine the most general solutions of the reflection equations. With these results we were able to determine integrable boundary conditions for an anyonic chain where the bulk Hamiltonian has $D(D_3)$ symmetry.  

The $R$-matrix associated with $D(D_3)$ is the simplest example in a heirarchy of solutions associated with $D(D_n)$ which solve  the Yang--Baxter equation \cite{FDIL2008}. All of these solutions are characterised by an absence of crossing unitarity, meaning that our general formalism is applicable on a wider scale.  

A future direction in this program of research  is to compute the Bethe ansatz solution of the open chain model derived above. Implementation of the algebraic Bethe ansatz appears problematic in this case, due to the lack of a suitable pseudovacuum state. A more promising avenue is offered by a functional approach aided by fusion relations e.g. see \cite{fnr2007} and references therein.

\end{document}